\documentclass[twoside,11pt]{article}

\usepackage{jmlr2e}
\usepackage{algorithm}
\usepackage[utf8]{inputenc} 
\usepackage[T1]{fontenc}    
\usepackage{hyperref}       
\usepackage{url}            
\usepackage{booktabs}       
\usepackage{amsfonts}       
\usepackage{nicefrac}       
\usepackage{amsmath,amssymb,graphicx}
\usepackage{color}
\usepackage{bm,bbm}
\usepackage{algorithm}
\usepackage{algpseudocode}
\usepackage[tight,footnotesize]{subfigure}

\begin{document}
	\title{Influential Node Detection in Implicit Social Networks using Multi-task Gaussian Copula Models}
	
	\author{\name Qunwei Li \email qli33@syr.edu \\
		\addr Syracuse University
		\AND
		\name Bhavya Kailkhura \email kailkhura1@llnl.gov \thanks{This work was supported in part by ARO under Grant W911NF-14-1-0339. This work was performed under the auspices of the U.S. Dept. of Energy by Lawrence Livermore National Laboratory under Contract DE-AC52-07NA27344.} \\
		\addr  Lawrence Livermore National Labs
		\AND
		\name Jayaraman J. Thiagarajan \email jjayaram@llnl.gov\\
		\addr Lawrence Livermore National Labs 
		\AND
		\name Zhenliang Zhang \email zhenliang.zhang@intel.com\\
		\addr Intel Labs
		\AND
		\name Pramod K. Varshney \email varshney@syr.edu\\
		\addr Syracuse University}
	
	\editor{Oren Anava, Marco Cuturi, Azadeh Khaleghi, Vitaly Kuznetsov, Alexander Rakhlin}
	
	\maketitle
	\begin{abstract}
		Influential node detection is a central research
		topic in social network analysis. Many existing
		methods rely on the assumption that the network structure
		is completely known \textit{a priori}. However, in
		many applications, network structure is unavailable
		to explain the underlying information
		diffusion phenomenon. To address the challenge of information diffusion analysis with incomplete knowledge of network structure,
		we develop a multi-task low rank linear influence model. By exploiting the relationships between contagions, our approach can simultaneously predict the volume (i.e. time series prediction) for each contagion (or topic) and automatically identify the most influential nodes for each contagion. The proposed model is validated using synthetic data and an ISIS twitter dataset. In addition to improving the volume prediction performance significantly, we show that the proposed approach can reliably infer the most influential users for specific contagions.
	\end{abstract}
	\section{Introduction}
	\label{sec:intro}
	Information emerges dynamically and diffuses quickly via agent interactions in complex networks (e.g. social networks) \citep{lopez2008diffusion}. Consequently, understanding and prediction of information diffusion mechanisms are challenging. There is a rapidly growing interest in exploiting knowledge of the information dynamics to better characterize the factors influencing spread of diseases, planned terrorist attacks, and effective social marketing campaigns, etc \citep{guille2012predictive}. The broad applicability of this problem in social network analysis has led to focused research on the following questions: (I) Which contagions are the most popular and can diffuse the most? (II) Which members of the network are influential and play important roles in the diffusion process? (III) What is the range over which the contagions can diffuse \citep{guille2013information}? While attempting to answer these questions, one is confronted with two crucial challenges. First, a descriptive diffusion model, which can mimic the behavior observed in real world data, is required. Second, efficient learning algorithms are required for inferring influence structure based on the assumed diffusion model. 
	
	A variety of information diffusion prediction frameworks have been developed in the literature \citep{yang2010modeling,wang2013sparse,guille2013information,du2013scalable,zhang2016towards}. A typical assumption in many of these approaches is that a connected network graph and knowledge of the corresponding structure are available {\it a priori}. However, in practice, the structure of the network can be implicit or difficult to model, e.g., modeling the structure of the spread of infectious disease is almost impossible. As a result, network structure unaware diffusion prediction models have gained interest. For example,  \citep{yang2010modeling}, Yang \textit{et. al.} proposed a linear influence model, which can effectively predict the information volume by assuming that each of the contagions spreads with the same influence in an implicit network. Subsequently, in \citep{wang2013sparse}, the authors extended LIM by exploiting the sparse structure in the influence function to identify the influential nodes. Though the relationships between multiple contagions can be used for more accurate modeling, most of the existing approaches ignore that information.
	
	In this paper, we address the above issues by augmenting linear influence models with complex task dependency information. More specifically, we consider the dependency of different contagions in the network, and characterize their relationships using Copula Theory. Furthermore, by imposing a low-rank regularizer, we are able to characterize the clustering structure of the contagions and the nodes in the network. Through this novel formulation, we attempt to both improve the accuracy of the prediction system and better regularize the influence structure learning problem. Finally, we develop an efficient algorithm based on proximal mappings to solve this optimization problem. Experiments with synthetic data reveal that the proposed approach fairs significantly better than a state-of-the-art multi-task variant of LIM both in terms of volume prediction and influence structure estimation performance. In addition, we demonstrate the superiority of the proposed method in predicting the time-varying volume of tweets using the ISIS twitter dataset\footnote{ISIS dataset from Kaggle is available at https://www.kaggle.com/kzaman/how-isis-uses-twitter.}. 
	
	\section{Background}
	\label{sec:background}
	In this section, we present the formulation of linear influence model (LIM) \citep{yang2010modeling} and discuss its limitations. Consider a set of $N$ nodes that participate in an information diffusion process of $K$ different contagions over time. Node $u \in \{1,\ldots,N\}$ can be infected by contagion $k\in\{1,\ldots,K\}$ at time $t\in\{0,1,\ldots,T\}$. The volume $V_k(t)$ is defined as the total number of nodes that get infected by the contagion $k$ at time $t$. Let the indicator function $M_{u,k}(t)=1$ represent the event that node $u$ got infected by contagion $k$ at time $t$, and $0$ otherwise. LIM models the volume $V_k(t)$ as a sum of influences of nodes $u$ that got infected before time $t$:
	\begin{align}
		V_k(t+1)=\sum_{u=1}^N\sum_{l=0}^{L-1}M_{u,k}(t-l)I_u(l+1),
		\label{lim_vk}
	\end{align}where each node $u$ has a particular non-negative influence function $I_u(l)$. One can simply think of $I_u(l)$ as the number of follow-up infections $l$ time units after $u$ got infected. The value of $L$ is set to indicate that the influence of a node drops to $0$ after $L$ time units. Thus, the influence of node $u$ is denoted by the vector $\mathbf I _u=\left(I_u\left(1\right),\ldots,I_u\left(L\right) \right)^T\in \mathbb R^{L\times 1} $. Next, using the notation $\mathbf V _k=\left(V(1),\ldots,V(T) \right)^T\in \mathbb R^{T\times 1} $ and $\mathbf{I} = (\mathbf I _1^T,\cdots,\mathbf I _N^T)^T\in \mathbb R^{LN\times 1} $, the inference procedure of LIM can be formulated as follows
	\begin{equation}\label{max}
		\text{minimize}\ \ \sum\limits_{k=1}^{K}\|{\mathbf{V}_k}-{\mathbf{M}_k}\cdot {\mathbf{I}}\|^2_2+ \mathbbm 1({\mathbf I}),
	\end{equation}where $\mathbf{M}_k$ is obtained via concatenation of $ M_{u,k}$, $\|\cdot\|_2$ denotes the Euclidean norm, and $\mathbbm 1({\mathbf I})$ is an indicator function that is zero when $I_{uk}\left(l\right)\ge 0$ and $+\infty$ otherwise. Though LIM has been effective in predicting the future volume for each contagion, it assumes that each node has the same influence across all the contagions. Consequently, to achieve contagion-sensitive node selection in an implicit network, the LIM model was extended and the multitask sparse linear influential model (MSLIM) was proposed in \citep{wang2013sparse}.
	
	The influence function is defined by extending ${\mathbf{I}}_u$ in LIM into contagion-sensitive ${\mathbf{I}}_{u,k}\in {\mathbb R}^{L\times 1}$, which is a $L$-length vector representing the influence of the node $u$  for the contagion $k$. For each contagion $k$, let ${\mathbf{I}}^k \in {\mathbb R}^{LN \times 1}$ be the vector obtained by concatenating ${\mathbf{I}}_{1k},\ldots,{\mathbf{I}}_{Nk}$. For each node $u$, the influence matrix for the node $u$ is defined: ${\mathbf{I}}_u=({\mathbf{I}}_{u1},\ldots,{\mathbf{I}}_{uK})\in {\mathbb R}^{L\times K}$. Using these notations, the inference procedure to estimate ${\mathbf{I}}_{u,k}$ was formulated as follows 
	\begin{equation}\label{m}
		\text{minimize}\ \ \frac{1}{2}\sum\limits_{k=1}^K\|{\mathbf{V}}_k-{\mathbf{M}}_k\cdot {\mathbf{I}}^k\|^2_2+\lambda\sum\limits_{u=1}^N\| {\mathbf I}_u\|_F+\gamma\sum\limits_{u=1}^N\sum\limits_{k=1}^K\| {\mathbf I}_{uk}\|_2+\mathbbm 1({\mathbf I}),
	\end{equation}where $\| \cdot\|_F$ denotes the Frobenius norm. The penalty term $\| {\mathbf I}_u\|_F$ was used to encourage the entire matrix $ {\mathbf I}_u$ to be zero altogether, which means that the node $u$ is non-influential for all different contagions. If the estimated $\| {\mathbf I}_u\|_F >0$ (i.e., the matrix $ {\mathbf I}_u$ is non-zero), a fine-grained selection is performed by the penalty $\sum\limits_{u=1}^N\sum\limits_{k=1}^K\| {\mathbf I}_{uk}\|_2$, which is essentially a group-Lasso penalty and can encourage the sparsity of vectors $\{ {\mathbf I}_{uk}\}$. For a specific contagion $k$, one can identify the most influential nodes by finding the optimal solution $\{\hat {\mathbf I}_{uk}\}$ of \eqref{m}.  However, the penalty terms used in MSLIM encourages that certain nodes have no influence over all the contagions which may not be true in practice. Furthermore, for most of the real world applications, there exists complex dependencies among the contagions. In order to alleviate these shortcomings, we propose a novel probabilistic multi-task learning framework and develop efficient optimization strategies.
	
	
	
	\section{Proposed Approach}
	\label{sec:approach}
	\noindent \textbf{Probabilistic Multi-Contagion Modeling of Diffusion: }We assume a linear regression model for each task:${\mathbf V}_k={\mathbf M}_k{\mathbf I}^k+{\bm {n}}_k$, where $\mathbf{V}_k,\mathbf{M}_k$ and $\mathbf{I}^k$ are defined as before, and $\mathbf n_k \in \mathbb R^{T\times 1}$ is an i.i.d. zero-mean Gaussian noise vector with the covariance matrix ${\bm \Sigma}{_k}$. The distribution for ${\mathbf V}_k$ given ${\mathbf M}_k$, ${\mathbf I}^k$ and ${\bm \Sigma}{_k}$ can be expressed as
	\begin{align}\label{prior}
		{\mathbf V}_k|{\mathbf M}_k, {\mathbf I}^k,  {\bm \Sigma}_{k}\sim{ \mathcal{N}}\left({\mathbf M}_k{\mathbf I}^k,{\bm \Sigma}_{k}\right)=\frac{\exp\left({-\frac{1}{2}}\left({\mathbf V}_k-{\mathbf M}_k{\mathbf I}^k\right)^T{\bm \Sigma}_{k}^{-1}\left({\mathbf V}_k-{\mathbf M}_k{\mathbf I}^k\right)\right)}{\left(2\pi\right)^{\frac{T}{2}}|{\bm \Sigma}_{k}|^{\frac{1}{2}}}.
	\end{align}Assuming that the influence for a single contagion is also Gaussian distributed, we can express the marginal distributions as ${\mathbf I}^k|{\mathbf m}_k,{\bm \Theta}_k \sim{\mathcal N}({\mathbf m}_k,{\bm \Theta}_k)$,
	{where ${\mathbf m}_k\in \mathbb R^{LN\times 1}$ is the mean vector and can be expressed as ${\mathbf m}_k=[{\mathbf m}_{1,k}^T,\ldots,{\mathbf m}_{N,k}^T]^T$, and ${\bm \Theta}_k \in \mathbb R^{LN\times LN}$ is the covariance matrix of $\mathbf I^k$. For a node $u$ and contagion $k$, we assume that the variables in the influence $\mathbf I_{uk}$ have the same mean, i.e., ${\mathbf m}_{u,k}=m_{u,k} \mathbf 1_{L\times1}$, where $m_{u,k}$ is a scalar and $\mathbf 1_{L\times1}$ is a vector of all ones with dimension ${L\times1}$. Let $\mathbf m^\prime \in \mathbb R^{N\times K}$ represent the mean matrix with entries $m_{u,k}$, and it is connected as ${\mathbf m}=\left({\mathbf m}_1,\ldots,{\mathbf m}_K\right)=\mathbf Q \mathbf m^\prime$, where $\mathbf Q\in \mathbb R^{LN\times N}=\mathcal I_{N\times N}\otimes \mathbf 1_{L\times 1}$ and $\mathcal I_{N\times N}$ is the identity matrix with dimension $N\times N$ and $\otimes$ is the Kronecker product operator.}
	
	\subsection{Dependence Structure Modeling Using Copulas}
	Consider a general case where the contagions are correlated. We construct a new influence matrix ${\mathbf I}=\left[{\mathbf I}^1,\ldots,{\mathbf I}^K \right]\in {\mathbb R}^{LN\times K}$.
	In our formulation, ${\mathbf I}^k$'s are assumed to be correlated and the joint distribution of ${\mathbf I}$ is not a simple product of all the marginal distributions of $\mathbf I^k$ as is adopted by most multi-task learning formulations.
	Here, we propose to use a multi-task copula that is obtained by tailoring the copula model for the multi-task learning problem. 
	\begin{theorem}
		(Sklar's Theorem). Consider an $N$-dimensional distribution function $F$ with marginal distribution functions $F_1,\ldots,F_N$. Then there exists a copula $C$, such that for all $x_1,\ldots,x_N$ in $[-\infty,\infty]$, $F(x_1,\ldots,x_N)=C\left( F_1\left(x_1 \right) ,\ldots, F_N\left(x_N \right) \right)$. If $F_n$ is continuous for $1\le n\le N$, then $C$ is unique, otherwise it is determined uniquely on $RanF_1\times \ldots \times RanF_N$ where $RanF_n$ is the range of $F_n$. Conversely, given a copula $C$ and univariate CDFs $F_1,\ldots, F_N$, $F$ is a valid multivariate CDF with marginals $F_1,\ldots, F_N$.
		
	\end{theorem}As a direct consequence of Sklar's Theorem, for continuous distributions, the joint probability density function (PDF) $f\left(x_1,\ldots,x_N \right) $ is obtained by,
	\begin{align}
		f(x_1,\ldots,x_N)=\left(\prod\limits_{n=1}^{N}f_n\left( x_n\right)  \right) c\left(F_1\left( X_1\right),\ldots,F_N\left( X_N\right)    \right), 
	\end{align}
	where $f_n\left( \cdot\right) $ is the marginal PDF and c is termed as the copula density given by 
	\begin{align}
		c(v)=\frac{\partial ^NC\left( v_1,\ldots,v_N\right) }{\partial v_1,\ldots,\partial v_N}
	\end{align}
	where $v_n=F_n(x_n)$. We extend the copula theory to multi-task learning and express the joint distribution of $\mathbf I$ as follows:
	\begin{align}
		p({\mathbf I}^1,{\mathbf I}^2,\ldots,{\mathbf I}^K)=\left(\prod _{k=1}^K{\mathcal N}({\mathbf m}_k,{\bm \Theta}_k)\right)c(F_1({\mathbf I}^1), F_2({\mathbf I}^2),\ldots,F_K({\mathbf I}^K)),
	\end{align}
	where $F_k({\mathbf I}^k)$ is the CDF of the influence for $k^{\text{th}}$ contagion. The copula density function $c(\cdot)$ takes all marginal CDFs $\{F_k({\mathbf I}^k) \}_{k=1}^K$ as its arguments, and maintains the output correlations in a parametric form.

	\noindent \textbf{Gaussian copula: }There are a finite number of well defined copula families that can characterize several dependence structures. Though, we can investigate the choice of an appropriate copula, we consider the Gaussian copula for its favorable analytical properties. A Gaussian copula can be constructed from the multivariate Gaussian CDF, and the resulting prior on $\mathbf I$ is given by a multivariate Gaussian distribution as
	\begin{align}\label{like}
		{\mathbf I}\sim {\mathcal {MN}}_{LN\times K} ({\mathbf m},\mathbf {\mathbf U}, {\mathbf \Omega})=\frac{\exp\left( -\frac{1}{2}\text{tr}\left( {\mathbf U}^{-1}\left( {\mathbf I}-\mathbf m \right) {\mathbf \Omega}^{-1}\left( {\mathbf I}-\mathbf m \right)^T\right)\right)}{\left( 2\pi \right)^{\frac{LNK}{2}}|{\mathbf \Omega}|^{\frac{LN}{2}}|{\mathbf U}|^{\frac{K}{2}}}
	\end{align}where ${\mathbf U}\in {\mathbb R}^{LN\times LN}$ is the row covariance matrix modeling the correlation between the influence of different nodes, ${\mathbf \Omega}\in {\mathbb R}^{K\times K}$ is the column covariance matrix modeling the correlation between the influence for different contagions, and $\mathbf m \in \mathbf R^{LN\times K}$ is the mean matrix of $\mathbf I$. The two covariances can be computed as $E\left[ \left( {\mathbf I}-\mathbf m \right)\left( {\mathbf I}-\mathbf m\right)^T\right]={\mathbf U}{\text{tr}}({\mathbf \Omega})$ and $E\left[ \left( {\mathbf I}-\mathbf m\right)^T\left( {\mathbf I}-\mathbf m\right)\right]={\mathbf \Omega}{\text{tr}}({\mathbf U})$ respectively. We assume that $N$ individual nodes are spreading the contagions and influencing others independently, and thus the row covariance matrix is diagonal and can be expressed as $
		{\mathbf U}=\text{diag}(e^2_1,e^2_2,\ldots,e^2_N)\otimes \mathcal I_{L\times L},
$
	where $e_n^2,n\in\{1,\ldots,N\}$ are scalars. The posterior distribution for $\mathbf I$, which is proportional to the product of the prior in Eq. \ref{prior} and the likelihood function in Eq. \ref{like}, is given as
	\begin{align}
		p\left({\mathbf I}|{\mathbf M}, {\mathbf V},{\bm \Sigma}, \mathbf U, {\bm \Omega}\right)&\propto p\left({\mathbf V}|{\mathbf M}, {\mathbf I},{\bm \Sigma}\right)p\left({\mathbf I}| \mathbf m, \mathbf U, {\bm \Omega}\right)\nonumber\\
		&=\left(\prod _{k=1}^K{ \mathcal{N}}\left({\mathbf M}_k{\mathbf I}^k,{\bm \Sigma}_{k}\right)\right){\mathcal {MN}}_{LN\times K}\left({\mathbf I}|\mathbf m,{\mathbf U},{\mathbf \Omega}\right),
	\end{align}where ${\mathbf M}=\left({\mathbf M}_1,\ldots , {\mathbf M}_K \right)\in {\mathbb R}^{T\times LNK} $, ${\mathbf V}=\left({\mathbf V}_1,\ldots , {\mathbf V}_K \right)\in {\mathbb R}^{T\times K}$, ${\bm \Sigma}$ is the corresponding covariance matrix of ${\bm n}=\left({\bm n}_1,\ldots , {\bm n}_K\right) \in {\mathbb R}^{T\times K} $. We assume ${\bm \Sigma}_{k}\triangleq \sigma^2 {\mathcal I}_{T\times T}$ and also an identical value of $e_n^2=e^2, \forall k=1,\ldots,K, \forall n=1,\ldots,N$. We employ maximum a posteriori (MAP) and maximum likelihood estimation (MLE), and obtain ${\mathbf I}$, $\mathbf m$, and $\mathbf \Omega$ by
	\begin{equation*}
		\begin{aligned}
			\underset{{\mathbf I},\mathbf m, \mathbf \Omega}{\text{min}}
			\frac{1}{\sigma^2} \sum\limits^K_{k=1} \| {\mathbf V}_k-{\mathbf M}_k{\mathbf I}^k\|_2^2 +\frac{1}{e^2}\text{tr}\left(({{\mathbf I}-\mathbf m}){\mathbf \Omega}^{-1}({\mathbf I}-\mathbf m)^T\right)+LN\ln|{\mathbf \Omega}|+\mathbbm 1({\mathbf I}).
		\end{aligned}
	\end{equation*}However, if we assume $\mathbf \Omega^{-1}$ to be non-sparse, the solution to $\mathbf \Omega^{-1}$ will not be defined (when $K>LN$) or will overfit (when $K$ is of the same order as $LN$) \citep{rai2012simultaneously}. In fact, some contagions in the network can be uncorrelated, which makes the corresponding entry values in $\mathbf \Omega^{-1}$ zero. Hence, we add a $l_1$ penalty to promote sparsity of matrix $\mathbf \Omega^{-1}$ to obtain
	\begin{equation*}
		\begin{aligned}
			\underset{{\mathbf I},\mathbf m, \mathbf\Omega}{\text{min}}
			\sum\limits^K_{k=1} \| {\mathbf V}_k-{\mathbf M}_k{\mathbf I}^k\|_2^2 +\lambda_1\text{tr}\left(({{\mathbf I}-\mathbf m}){\mathbf \Omega}^{-1}({\mathbf I}-\mathbf m)^T\right)-\lambda_2\ln|{\mathbf \Omega}^{-1}|+\lambda_3\| {{\mathbf \Omega}}\|_{1}+\mathbbm 1({\mathbf I}).
		\end{aligned}
	\end{equation*}
	\subsection{Modeling Structure of Influence Matrix $\mathbf I$}
	In order to better characterize the influence matrix, we propose to impose a low rank structure on the influence matrix $\mathbf{I}$. The nodes or the contagions in the influence network are known to form communities (or clustering structures), which may be captured using the low-rank property of the influence matrix. Note that, the sparse structure in the influence matrix implies that most individuals only influence a small fraction of contagions in the network while there can be a few nodes with wide-spread influence. We incorporate this into our formulation by using a sparsity promoting regularizer over $\mathbf I_{u,k}$.
	\begin{equation}\label{opt}
		\begin{aligned}
			& \underset{{\mathbf I},\mathbf m, \mathbf \Omega}{\text{min}}
			& &  \sum\limits^K_{k=1} \| {\mathbf V}_k-{\mathbf M}_k{\mathbf I}^k\|_2^2 +\lambda_1\text{tr}\left(({{\mathbf I}-\mathbf m}){\mathbf \Omega}^{-1}({\mathbf I}-\mathbf m)^T\right)\\
			& & &-\lambda_2\ln|{\mathbf \Omega}^{-1}|+\lambda_3\| {{\mathbf \Omega}}\|_{1}+\lambda_4\| {\mathbf I}\|_{\ast}+\lambda_5\sum\limits_{u=1}^N\sum\limits_{k=1}^K\|{\mathbf I}_{uk}\|_2+\mathbbm 1({\mathbf I}),
		\end{aligned}
	\end{equation}where $\|\cdot\|_{\ast}$ denotes the nuclear norm, and $\lambda_1$, $\lambda_2$, $\lambda_3$, $\lambda_4$ and $\lambda_5$ are the regularization parameters. With the estimated $\{\hat {\mathbf I}_{uk}\}$, one can predict the total volume of the contagion $k$ at $T+1$ by $\hat V_k(T+1)=\sum_{u=1}^N\sum_{l=0}^{L-1}M_{uk}(T-l)I_{uk}(l+1)$. 
	

	\section{Algorithm}
	\label{sec:algorithm}
	We adopt an alternating optimization approach to solve the problem in Eq. \ref{opt}.
	
	\noindent \textbf{Optimization w.r.t. $\mathbf m$: }Given ${\mathbf I}$ and $\mathbf \Omega^{-1}$, the mean matrix $\mathbf m$ can be obtained by solving the following problem
	\begin{equation*}
		\begin{aligned}
			& \underset{\mathbf m}{\text{min}}
			& & \text{tr}\left(({{\mathbf I}-\mathbf m}){\mathbf \Omega}^{-1}({\mathbf I}-\mathbf m)^T\right).\\\end{aligned}
	\end{equation*}
	The estimate  $\hat {\mathbf m}$ can be analytically obtained as $\hat {\mathbf m}=\frac 1 L \mathbf Q\mathbf Q^T \mathbf I$.
	
	\noindent \textbf{Optimization w.r.t. $\mathbf \Omega$: }Given ${\mathbf I}$ and $\mathbf m$, the contagion inverse covariance matrix $\mathbf \Omega ^{-1}$ can be estimated by solving the following optimization problem
	\begin{equation*}
		\begin{aligned}
			& \underset{\mathbf\Omega}{\text{min}}
			& &  \lambda_1\text{tr}\left(({{\mathbf I}-\mathbf m}){\mathbf \Omega}^{-1}({\mathbf I}-\mathbf m)^T\right)-\lambda_2\ln|{\mathbf \Omega}^{-1}|+\lambda_3\| {{\mathbf \Omega}}\|_{1}
		\end{aligned}
	\end{equation*}The above is an instance of the standard inverse covariance estimation problem with sample covariance $\frac{\lambda_1}{\lambda_2}({\mathbf I}-\mathbf m)^T({\mathbf I}-\mathbf m)$, which can be solved using standard tools. In particular, we use the graphical Lasso procedure in \citep{friedman2008sparse}
	\begin{align}
		\hat {\mathbf \Omega}^{-1}=gLasso \left(\lambda_1/\lambda_2({\mathbf I}-\mathbf m)^T({\mathbf I}-\mathbf m),\lambda_3\right).
	\end{align}
	
	\noindent \textbf{Optimization w.r.t. ${\mathbf I}$: }The corresponding optimization problem becomes
	\begin{equation*}
		\begin{aligned}
			\underset{{\mathbf I}}{\text{min}}
			\sum\limits^K_{k=1} \| {\mathbf V}_k-{\mathbf M}_k{\mathbf I}^k\|_2^2 +\lambda_1\text{tr}\left(({{\mathbf I}-\mathbf m}){\mathbf \Omega}^{-1}({\mathbf I}-\mathbf m)^T\right)+\lambda_4\| {\mathbf I}\|_{\ast}+\lambda_5\sum\limits_{u=1}^N\sum\limits_{k=1}^K\|{\mathbf I}_{uk}\|_2+\mathbbm 1({\mathbf I}).
		\end{aligned}
	\end{equation*}We rewrite the problem as
	\begin{equation}\label{Iopt}
		\begin{aligned}
			& \underset{{\mathbf I}}{\text{min}}
			& &  \ell({\mathbf I})+\lambda_4\| {\mathbf I}\|_{\ast}+{\mathbbm{1}}({\mathbf I}).
		\end{aligned}
	\end{equation}
	where $\ell({\mathbf I})=\sum\limits^K_{k=1} \| {\mathbf V}_k-{\mathbf M}_k{\mathbf I}^k\|_2^2 +\lambda_1\text{tr}\left(({{\mathbf I}-\mathbf m})\mathbf \Omega^{-1}({\mathbf I}-\mathbf m)^T\right)+\lambda_5\sum\limits_{u=1}^N\sum\limits_{k=1}^K\|{\mathbf I}_{uk}\|_2$. This formulation involves a sum of a convex differentiable loss and convex non-differentiable regularizers which renders the problem non-trivial. A string of algorithms have been developed for the case where the optimal solution is easy to compute when each regularizer is considered in isolation. This corresponds to the case where the proximal operator defined for a convex regularizer $R: \mathbb R^{LN\times K}\rightarrow \mathbb R$ at a point $\mathbf Z$ by
	$
		\text{prox} _R(\mathbf Z)={\arg \min} \frac{1}{2}\|\mathbf I-\mathbf Z\|_F^2+R(\mathbf I),
	$
	is easy to compute for each regularizer taken separately. See \citep{combettes2011proximal} for a broad overview of proximal methods. The proximal operator for the nuclear norm is given by the shrinkage operation as follows \citep{beck2009fast}. If ${U}\text{diag} (\sigma_1,\ldots,\sigma_n)V^T$ is the singular value decomposition of $\mathbf Z$, then
	$
		\text{prox} _{\lambda_4\|\cdot\|_{\ast}}(\mathbf Z)={U}\text{diag} ((\sigma_i-\lambda_4)_{+})_iV^T
	$. The proximal operator of the indicator function ${\mathbbm{1}}({\mathbf I})$ is simply the projection onto $ I_{u,k}(l)\ge0$, which is denoted by $P_{\mathbbm{1}}({\mathbf I})$. Next, we mention a matching serial algorithm introduced in \citep{bertsekas2011incremental}. We present here a version where updates are performed according to a cyclic order \citep{richard2012estimation}. Note that one can also randomly select the order of the updates. We use the optimization algorithm~\ref{array-sum} to solve the optimization problem in Eq. \ref{Iopt}.
	
	\begin{algorithm}[t]
		\caption{Incremental Proximal Descent}
		\label{array-sum}
		\begin{algorithmic}[1]
			\State \text{Initialize} $\mathbf I=\mathbf A$
			\Repeat
			\State{Set $\mathbf I=\mathbf I-\theta \nabla_{\mathbf I}\ell({\mathbf I})$}
			\State{Set $\mathbf I=\text{prox} _{\theta\lambda_4\|\cdot\|_{\ast}}(\mathbf I)$}
			\State{Set $\mathbf I=P_{\mathbbm{1}}({\mathbf I})$}
			\Until{convergence}\\
			\Return $\mathbf I$
		\end{algorithmic}
	\end{algorithm}

	\section{Experiments}
	\label{sec:results}
	We compare the performance of the proposed approach to MSLIM by applying it to both synthetic and real datasets. Since the volume of a contagion over time $V_k(t)$ can be viewed as a time series, we set up this problem as a time series prediction task and evaluate the performance using the prediction mean-squared error (MSE). Furthermore, for the synthetic data set, where we have access to the true influence matrix $\mathbf{I}$, we also evaluate the performance of the influence matrix prediction task using the metric $\|\hat{\mathbf{I}}-\mathbf{I}\|_F$. We determined the regularization parameters for the proposed model using cross validation. In particular, we split the first $60\%$ of the time instances as the training set and the rest for validation. Following~\citep{wang2013sparse}, we combine the training and validation sets to re-train the model with the best selected regularization parameters and estimate the influence matrix. 
	
	\subsection{Synthetic Data}
	We created a synthetic dataset with the number of nodes fixed at $N=100$ and the number of contagions at $K=20$. In addition, we assumed that $L=10$ and $T=20$. A rank $5$ (low-rank) influence matrix $\mathbf{I}$ was generated randomly with uniformly distributed entries.
	The matrix $\mathbf M$ was generated with uniformly distributed random integers $\{0,1\}$. 
	Following our model assumption, the volume for each $\mathbf{V}_k$ was calculated as follows $\mathbf{V}_k = \mathbf{M}_k\times \mathbf{I}^k + \mathcal{N}(\mathbf{0},\mathbf{\Delta})$ where $\mathcal{N}(\mathbf{0},\mathbf{\Delta})$ is a multivariate normal distribution with covariance matrix $\Delta$. In Table~\ref{synthetic}, we present the results obtained using the proposed approach and its comparison to MSLIM. As can be observed, for both volume prediction and influence matrix estimation tasks, the proposed approach achieves highly accurate estimates. 
	
	\begin{table} 
		\caption{Prediction performance for different information diffusion models on synthetic data.}
		\begin{center}
			\renewcommand{\arraystretch}{1.2}
			\begin{tabular}{ | c | c | c |}
				\hline
				\textbf{Approach}& MSLIM & Proposed \\ \hline
				\textbf{Volume Prediction MSE} & 0.834 & \textbf{0.007} \\ \hline \hline
				\textbf{Influence Matrix Estimation Error} & 0.7681 & \textbf{0.62} \\
				\hline
			\end{tabular}
			\label{synthetic}
		\end{center}
	\end{table}
	
	\subsection{ISIS Twitter Data}
	In this section, we demonstrate the application of the proposed approach to a real-word analysis task. We begin by describing the twitter dataset used for analysis and the procedure adopted to extract the set of contagions. Following this, we discuss the problem setup and present comparisons to MSLIM on predicting the time-varying tweet volume. Finally, we present a qualitative analysis of the inferred influence structure for different contagions.
	
	The ISIS dataset from Kaggle\footnote{ISIS dataset from Kaggle is available at https://www.kaggle.com/kzaman/how-isis-uses-twitter.} is comprised of over $17,000$ tweets from $112$ users posted between January 2015 and May 2016. In addition to the actual tweets, meta-information such as the user name and the timestamp for each tweet are included. We performed a standard pre-processing by removing a variety of stop words, e.g. URLs, symbols. After preprocessing, we converted each tweet into a bag-of-words representation and extracted the term frequency-inverse document frequency (tf-idf) feature. 
	
	\noindent \textbf{Topic Modeling: }When applying our approach, the first step is to define semantically
	meaningful contagions. A simple way of defining topics is to directly
	use words as topics (e.g., ISIS). However, a single
	word may not be rich enough to represent a broad topic
	(e.g., social network sites). Hence, we propose to perform topic modeling on the tweets based on the tf-idf features. In our experiment, we obtained the topics using Non-negative Matrix Factorization (NMF), which is a popular scheme for topic discovery, with the number
	of topics $K$ set at $10$. Table~\ref{BOW} lists the top $10$ words for each of the topics learned using NMF.
	
	
	\begin{table} 
		\caption{Top words for each topic learned using NMF with the ISIS twitter dataset.}
		\begin{center}
			\renewcommand{\arraystretch}{1.2}
			\begin{tabular}{ | l | c |}
				\hline
				\small
				\textbf{Topic 1} & isis ramiallolah iraq attack libya warreporter1 saa aamaq usa abu  \\ \hline 
				\textbf{Topic 2} & killed soldiers today airstrikes injured wounded civilians militants iraqi attack  \\ \hline
				\textbf{Topic 3} & syria russia ramiallolah turkey ypg breakingnews usa group saa terror  \\ \hline 
				\textbf{Topic 4} & state islamic fighters fighting group saudi new http wilaya control  \\ \hline
				\textbf{Topic 5} & aleppo nid gazaui rebels north today northern syrian ypg turkish  \\ \hline
				\textbf{Topic 6} & assad regime myra forces rebels fsa pro islam syrian jaysh  \\ \hline 
				\textbf{Topic 7} & al qaeda nusra abu sham ahrar islam jabhat http warreporter1  \\ \hline
				\textbf{Topic 8} & army iraq near ramiallolah iraqi lujah turkey ramadi west sinai  \\ \hline 
				\textbf{Topic 9} & allah people muslims abu accept muslim make know don islam  \\ \hline
				\textbf{Topic 10}& breaking islamicstate forces amaqagency city fighters iraqi near area syrian \\ \hline  
			\end{tabular}
			\label{BOW}
		\end{center}
	\end{table}
	
	\begin{figure}[th] 
		\centering
		\subfigure[Average Influence]{\includegraphics[width=.4\linewidth]{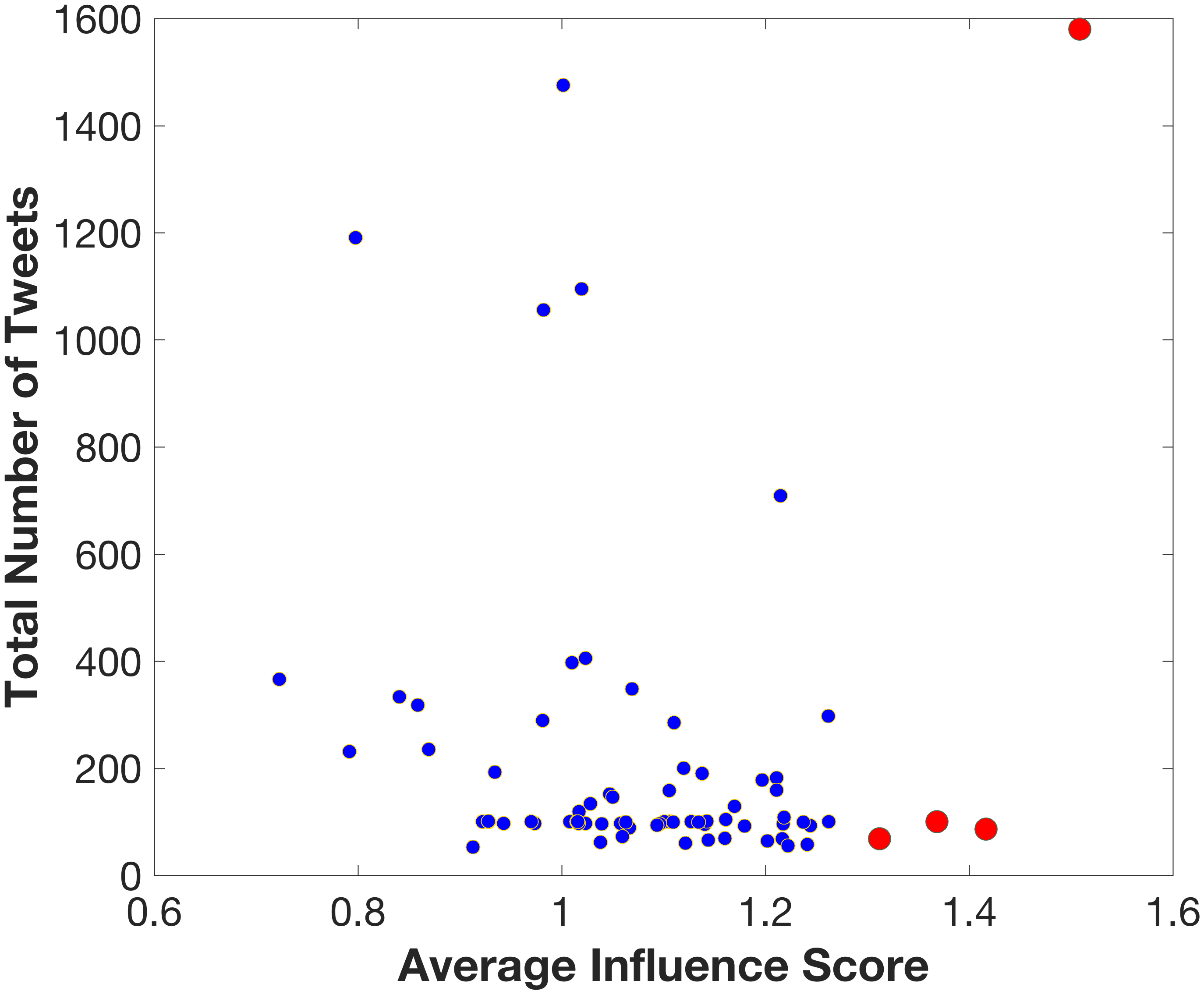}\label{fig:avg}}\hfill
		\subfigure[Maximum Influence]{\includegraphics[width=.4\linewidth]{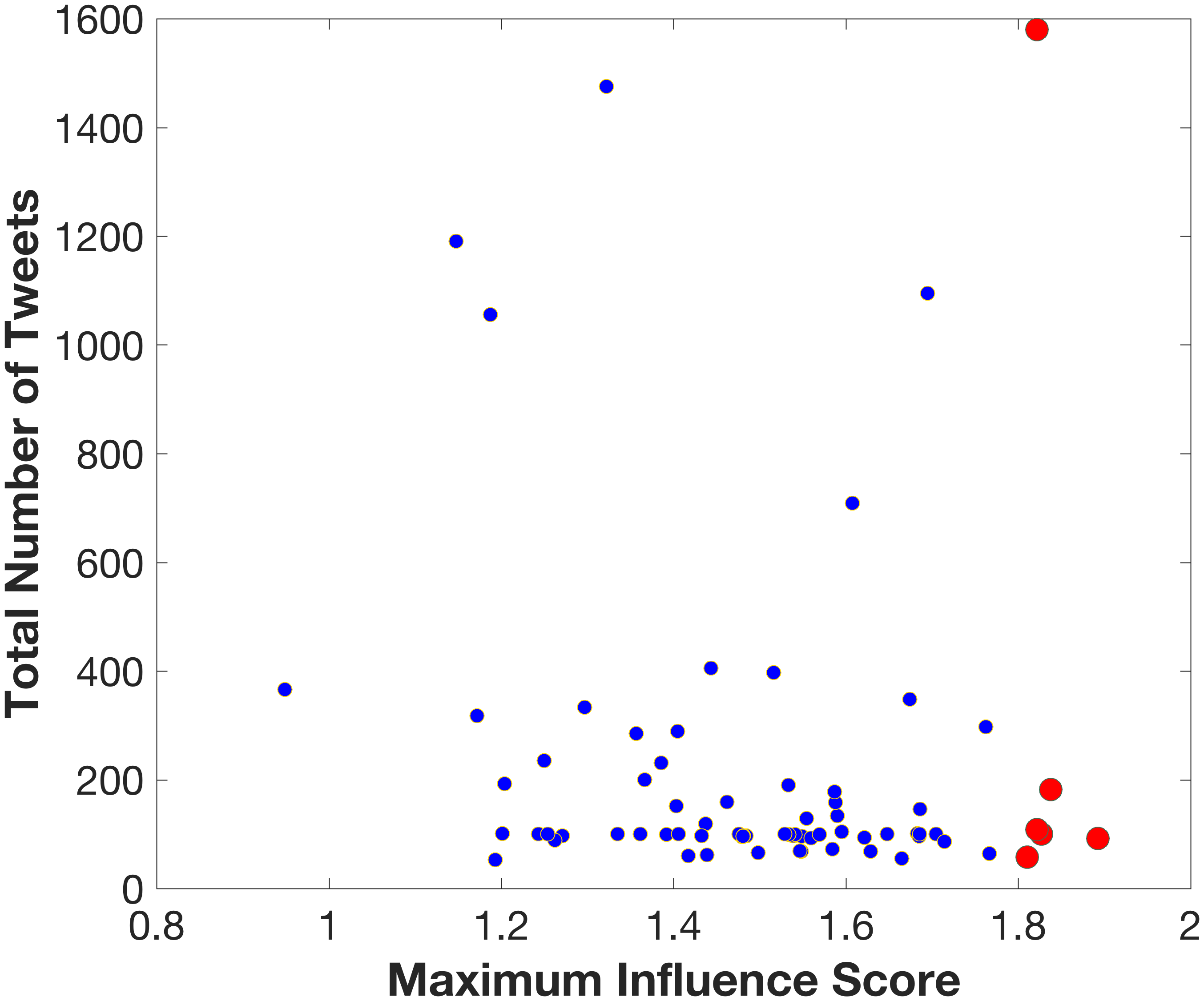}\label{fig:max}}
		
		\caption{Comparing statistics from the estimated influence matrix with the volume of tweets corresponding to each of the users to identify influential users. In both cases, the users with a large influence score are marked in red.}
	\end{figure}
	
	\begin{figure}[th] 
		\centering
		\subfigure[]{\includegraphics[width=.38\linewidth]{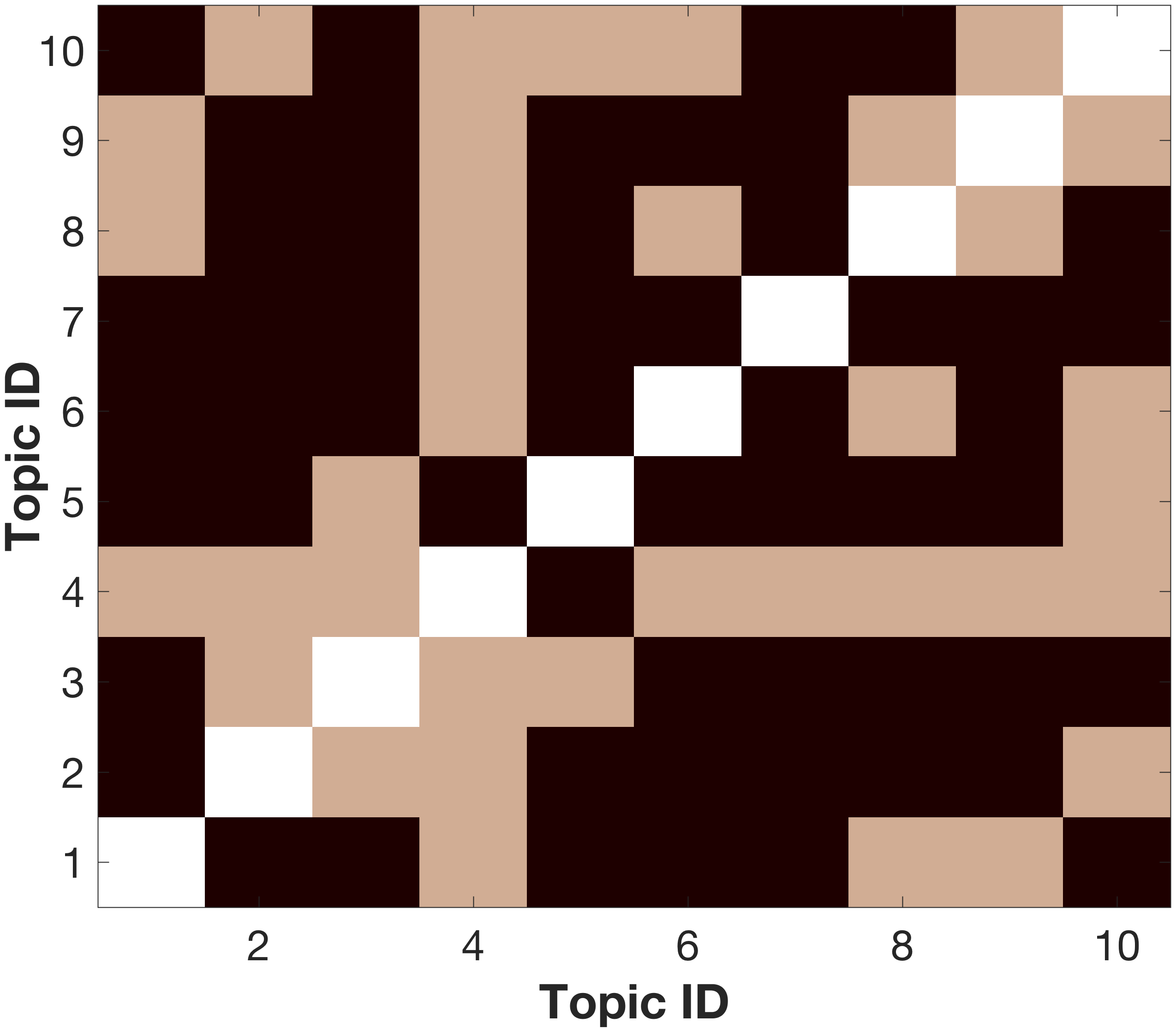} \label{fig:cov}}\hfill
		\subfigure[]{\includegraphics[width=.4\linewidth]{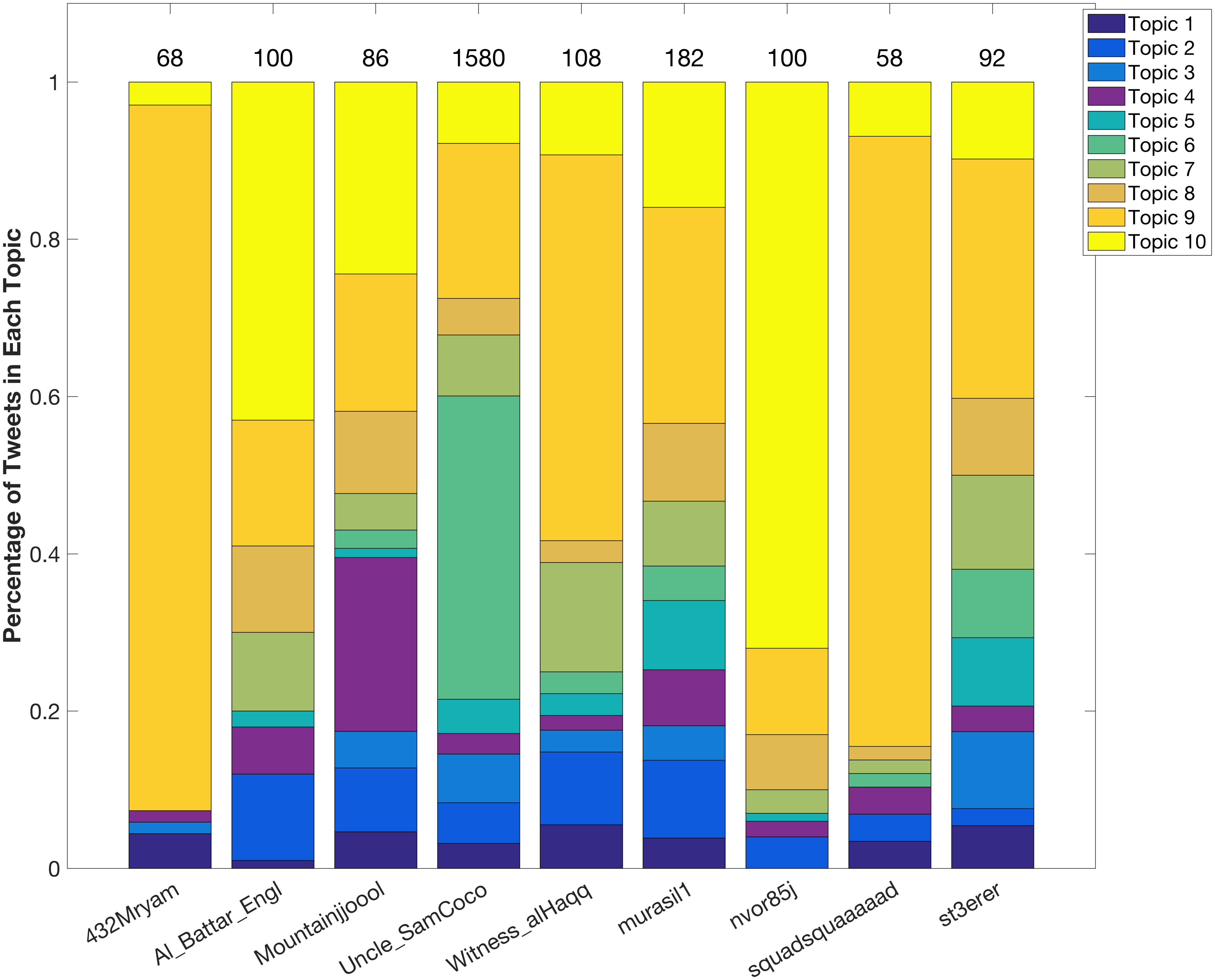}\label{fig:inf}}
		
		\caption{(a) Correlation Structure among the topics (non-black color represents positive correlation), (b) Top 9 influential users and their tweet distributions.}
	\end{figure}

	\noindent \textbf{Volume Time Series Prediction: }In our experiment, we
	set one day as the discrete time step for aggregating the tweet volume. The parameter $L$ denotes the number of time steps it takes for the influence of a user to decay to zero. We set the parameter $L$ equal to $5$
	since we observed that beyond $L = 5$, there is hardly any improvement in performance. The MSE on the predicted volume is computed over the entire period of observation. The comparison of the prediction MSE is presented in Table \ref{real}. It can be seen that the proposed approach significantly outperforms MSLIM in predicting the time-varying volume.
	
	\noindent \textbf{Influential Node Detection: }For a contagion $k$, we identify the most influential nodes with respect to this contagion as nodes having high $\|\mathbf{I_{u,k}}\|_2$ values. First, in Figure~\ref{fig:cov}, we plot the correlation among $10$ topics learned by NMF. More specifically, we plot the pair-wise correlation structure learned by our approach. It can be seen that, a strong positive correlation structure exists, which enabled the improved prediction in Table \ref{real}. Following this, we use the predicted influence matrix to select a set of highly influential nodes from the dataset. A simple approach to select the influential users can be to select the ones with a large number of tweets. However, we argue that the influence predicted in an information diffusion model can be vastly different. Consequently, we consider a user to be influential if she has a high influence score for at least one of the topics, or if she can be influential for multiple topics. For example, in Figure~\ref{fig:avg}, we plot average influence scores of the users (averaged over all the topics) against the total number of tweets. Similarly, in Figure~\ref{fig:max}, we plot influence scores of the users (maximum over all the topics) against the total number of tweets. The first striking observation is that the users with high influence scores are not necessarily the ones with the most number of tweets. Instead, their impact on the information diffusion relies heavily on the complex dynamics of the implicit network. 
	
	\begin{table}[t] 
		\centering
		\caption{Volume prediction performance on the ISIS twitter dataset.}
		\renewcommand{\arraystretch}{1.2}
		\begin{tabular}{| c | c | c |}
			\hline
			\textbf{Approach} & MSLIM & Proposed \\ \hline
			\textbf{Volume Prediction MSE} & 2.7 & \textbf{0.329} \\ \hline
		\end{tabular}
		\label{real}
	\end{table}

	Finally, in Figure~\ref{fig:inf} we plot the percentage of tweets regarding each of the topics for top $9$ influential nodes. Influential nodes are obtained as a union of nodes identified based on both average and maximum influence scores. More specifically, we select the union of users with average influence score greater than $1.3$ and maximum influence score greater than $1.8$. In addition to displaying the distribution across topics, for each influential user, we show the total number of tweets posted by that user. It can be seen that the total number of tweets of these users vary a lot and, therefore, is not a good indication of their influence.

	\section{Conclusion}
	In this paper, we considered the problem of influential node detection and volume time series prediction. We proposed a descriptive diffusion model to take dependencies among the topics into account. We also proposed an efficient algorithm based on alternating methods to perform inference and learning on the model. It was shown that the proposed technique outperforms existing influential node detection techniques. 
	Furthermore, the proposed model was validated both on a synthetic and a real (ISIS) dataset.
	We showed that the proposed approach can efficiently select the most influential users for specific contagions. We also presented several
	interesting patterns of the selected influential users for the ISIS dataset.
	
	
	\newpage
	
	\pagebreak

	\bibliography{ref_Lqw}

\begin{thebibliography}{13}
\providecommand{\natexlab}[1]{#1}
\providecommand{\url}[1]{\texttt{#1}}
\expandafter\ifx\csname urlstyle\endcsname\relax
  \providecommand{\doi}[1]{doi: #1}\else
  \providecommand{\doi}{doi: \begingroup \urlstyle{rm}\Url}\fi

\bibitem[Beck and Teboulle(2009)]{beck2009fast}
Amir Beck and Marc Teboulle.
\newblock A fast iterative shrinkage-thresholding algorithm for linear inverse
  problems.
\newblock \emph{SIAM journal on imaging sciences}, 2\penalty0 (1):\penalty0
  183--202, 2009.

\bibitem[Bertsekas(2011)]{bertsekas2011incremental}
Dimitri~P Bertsekas.
\newblock Incremental gradient, subgradient, and proximal methods for convex
  optimization: A survey.
\newblock \emph{Optimization for Machine Learning}, 2010:\penalty0 1--38, 2011.

\bibitem[Combettes and Pesquet(2011)]{combettes2011proximal}
Patrick~L Combettes and Jean-Christophe Pesquet.
\newblock Proximal splitting methods in signal processing.
\newblock In \emph{Fixed-point algorithms for inverse problems in science and
  engineering}, pages 185--212. Springer, 2011.

\bibitem[Du et~al.(2013)Du, Song, Gomez-Rodriguez, and Zha]{du2013scalable}
Nan Du, Le~Song, Manuel Gomez-Rodriguez, and Hongyuan Zha.
\newblock Scalable influence estimation in continuous-time diffusion networks.
\newblock In \emph{Advances in neural information processing systems}, pages
  3147--3155, 2013.

\bibitem[Friedman et~al.(2008)Friedman, Hastie, and
  Tibshirani]{friedman2008sparse}
Jerome Friedman, Trevor Hastie, and Robert Tibshirani.
\newblock Sparse inverse covariance estimation with the graphical lasso.
\newblock \emph{Biostatistics}, 9\penalty0 (3):\penalty0 432--441, 2008.

\bibitem[Guille and Hacid(2012)]{guille2012predictive}
Adrien Guille and Hakim Hacid.
\newblock A predictive model for the temporal dynamics of information diffusion
  in online social networks.
\newblock In \emph{Proceedings of the 21st international conference on World
  Wide Web}, pages 1145--1152. ACM, 2012.

\bibitem[Guille et~al.(2013)Guille, Hacid, Favre, and
  Zighed]{guille2013information}
Adrien Guille, Hakim Hacid, Cecile Favre, and Djamel~A Zighed.
\newblock Information diffusion in online social networks: A survey.
\newblock \emph{ACM SIGMOD Record}, 42\penalty0 (2):\penalty0 17--28, 2013.

\bibitem[L{\'o}pez-Pintado(2008)]{lopez2008diffusion}
Dunia L{\'o}pez-Pintado.
\newblock Diffusion in complex social networks.
\newblock \emph{Games and Economic Behavior}, 62\penalty0 (2):\penalty0
  573--590, 2008.

\bibitem[Rai et~al.(2012)Rai, Kumar, and Daume]{rai2012simultaneously}
Piyush Rai, Abhishek Kumar, and Hal Daume.
\newblock Simultaneously leveraging output and task structures for
  multiple-output regression.
\newblock In \emph{Advances in Neural Information Processing Systems (NIPS)},
  pages 3185--3193, 2012.

\bibitem[Richard et~al.(2012)Richard, Savalle, and
  Vayatis]{richard2012estimation}
Emile Richard, Pierre-andre Savalle, and Nicolas Vayatis.
\newblock Estimation of simultaneously sparse and low rank matrices.
\newblock In \emph{Proceedings of the 29th International Conference on Machine
  Learning (ICML)}, pages 1351--1358, 2012.

\bibitem[Wang et~al.(2013)Wang, Xiang, and Chang]{wang2013sparse}
Yingze Wang, Guang Xiang, and Shi-Kuo Chang.
\newblock Sparse multi-task learning for detecting influential nodes in an
  implicit diffusion network.
\newblock In \emph{AAAI}, 2013.

\bibitem[Yang and Leskovec(2010)]{yang2010modeling}
Jaewon Yang and Jure Leskovec.
\newblock Modeling information diffusion in implicit networks.
\newblock In \emph{2010 IEEE International Conference on Data Mining}, pages
  599--608. IEEE, 2010.

\bibitem[Zhang et~al.(2016)Zhang, He, Long, Huang, and Zhang]{zhang2016towards}
Peng Zhang, Jing He, Guodong Long, Guangyan Huang, and Chengqi Zhang.
\newblock Towards anomalous diffusion sources detection in a large network.
\newblock \emph{ACM Transactions on Internet Technology (TOIT)}, 16\penalty0
  (1):\penalty0 2, 2016.

\end{thebibliography}
	
\end{document}